\documentclass[11pt,twoside]{article}

\usepackage{asp2006}
\usepackage{epsf}
\usepackage{psfig}
\usepackage{lscape}

\begin{document}

  \title{ The jet of Markarian 501 from the sub-parsec to the kpc scale} 
\author{Gabriele Giovannini\altaffilmark{1,2}, 
Marcello Giroletti\altaffilmark{1}, 
Greg B. Taylor\altaffilmark{3}}
\altaffiltext{1}{Istituto di Radioastronomia-INAF, via Gobetti 101, 40129
Bologna, (I); m.giroletti@ira.inaf.it}
\altaffiltext{2}{Dipartimento di Astronomia, Universita' di Bologna, 
via Ranzani 1, 40127 Bologna, (I);
gabriele.giovannini@unibo.it}
\altaffiltext{3}{Department of Physics and Astronomy, University of New Mexico
(USA);gbtaylor@unm.edu}

\begin{abstract} 
We have observed the BL Lac object 
Markarian 501
at 1.4 GHz with the High Sensitivity Array and at 86 GHz with the global
VLBI mm array.
Thanks to the great resolution and sensitivity provided by these instruments,
we probe regions of the radio jets never accessed before.
The new data at 1.4 GHz allow us to map the one-sided jet at large distances
from the core, and to constrain jet properties thanks to the high jet to
counter-jet brightness ratio.
The 86 GHz data give us a high resolution image of the nuclear region.
Putting together these new results and available published data we
discuss the properties of this source from sub-parsec to kiloparsec
scales.
\end{abstract}

\section{Introduction}  

The study of extragalactic radio jets is an important area in
astrophysics. In radio loud sources, jets contribute a large fraction of the
total radiated power, and provide energy to kiloparsec scale radio
lobes.
Jets are
present in high and low power sources, with some common features: on the
parsec scale they are
relativistic regardless of power, and they are also intrinsically identical 
in beamed
and misaligned sources. \citet{gio01}, have shown that the Lorentz 
factor in
the parsec scale jets of low power FRI radio galaxies as well as of more
powerful FR IIs are both in the range $\Gamma = 3 - 10$. \citet{gir04b}
have
also shown that with these properties one can unify BL Lac objects and FRI
radio galaxies, if the former have jet axis oriented at an average viewing
angle of $\langle \theta \rangle = 18^\circ \pm 5^\circ$.

In this work we focus on the jet structure of the BL Lac source
Markarian 501. 
This object is highly active and well-studied at all frequencies. In the
radio band, centimeter VLBI observations have revealed a
clear limb-brightened structure, beginning in the very inner jet, suggestive of
a dual velocity structure \citep{gir04a}.
The complex
limb-brightened structure makes component identification problematic and
multi-epoch attempts to measure pattern speed conclude that it is not well
defined \citep{gir04a} or in any case at most subliminal \citep{edw02}. 
These seem to be common features in TeV blazars
\citep{pin04,gir06}, and theoretical models have 
been proposed to reconcile
the low observed speeds with the very high energy emission observed 
\citep{ghi05,wan04}.

However, the information available up to now has been restricted to the region
between $\sim 1$ and $\sim 100$ milliarcseconds, with smaller and larger scale
regions being precluded by inadequate resolution and sensitivity.
Improvements in the technical and organizational issues are now offering to
astronomers VLBI arrays of unprecedented resolution and sensitivity, such as
the High Sensitivity Array (HSA, see http://www.nrao.edu/hsa), and
the Global mm-VLBI Array 
(GMVA see http://www.mpifr-bonn.mpg.de/div/vlbi/globalmm).  
Thanks to its
proximity and brightness, Mrk 501 is an ideal laboratory for experiments using
these advanced VLBI techniques: it is at $z = 0.034$ (1 mas = 0.67 pc, using
$H_0 = 70$ km s$^{-1}$ Mpc$^{-1}$). The total flux density at 5 GHz is $S_5 =
1.4$ Jy; the Schwarzschild radius for its central black hole is estimated
about $10^{-4}$ pc (1.4 $10^{-4}$ mas), if we adopt
$M_\mathrm{BH} = 10^9 M_\odot$ \citep{rie03}.
Using these new facilities, we now access region 
never studied
previously: the jet base with the GMVA, and the faint, resolved jet region at
$>100$ mas with HSA.

\section {Observations}

We observed Mrk 501 on November 2004 with the HSA for 8 
hours at 1.4 GHz. 
Thanks to the HSA facility we observed with the full VLBA (NRAO), the VLA
phased array (NRAO), the Effelsberg (D) 100 m
single dish and the Green Bank Telescope (NRAO). The final map thanks to the 
high sensitive and good
uv-coverage, allows us to image the one sided low brightness jet at a 
much larger
distance (250 mas) from the core than previous observations \citep{gir04a},
and to obtain significant upper limits to the jet brightness ratio because of
the low noise level (0.05 mJy/beam with a HPBW = 9 $\times$ 5 mas).

We used the VLA phased array observations to produce also a VLA only image. 
During the 
observations the VLA was in the A configuration and we observed 3C 286 for
absolute flux calibration. 
At the VLA angular resolution Mrk 501 was strong and 
unresolved therefore it was used to calibrate the data and to phase the array.

On 14 Oct 2005 we observed Mrk 501 with the Global mm-VLBI Array (GMVA). 
The standard
frequency was 86.198 GHz, with a bandwidth of 128 MHz divided in eight 16 MHz
IFs. The participating telescopes were Effelsberg, Pico Veleta, the Plateau de
Bure interferometer, Onsala, Mets\"ahovi, and 8 VLBA stations (i.e.\ all except
Saint Croix and Hancock). The European telescopes observed for $\sim 9$ hours
and the American ones joined in for the last $\sim 6$ hours (Mauna Kea only for
the last $\sim 4$ hours). This experiment tested the sensitivity limits of the
array, since on the basis of the observed centimeter wavelength flux density
and spectral index \citep{gir04a}, Mrk 501 was expected to be only a few
hundreds of milliJanskys at this frequency.
The calibrator 3C345 was well detected on all baselines, except those to
Mets\"ahovi and North Liberty. From the fringe fitting of 3C345 we determined
rates and single-band
delays, and applied them to the whole data set. We obtained an image
of 3C345 and found it to be in agreement with published images of comparable or
slightly lower resolution \citep{lob00,lis05}.
At this stage, it was then possible to fringe fit Mrk 501 itself, averaging
over the 8 IFs. A solution interval as long as the scan, and a SNR cutoff 
of 3.0 was used. 

\section {Arcsecond scale}

Mrk 501 was observed with the VLA by \citet{koo92} and by \citet{cas99}. 
In these images the source is core dominated, but 
a two sided extended structure is also visible, which has been 
identified as the symmetric extended
emission characteristic of a radio galaxy. The source 
structure is in agreement with
an orientation near to the line of sight as expected from a BL Lac source.
The symmetric emission
implies that at this distance from the core no relativistic jet
remains.

\begin{figure}[!ht]

\plottwo{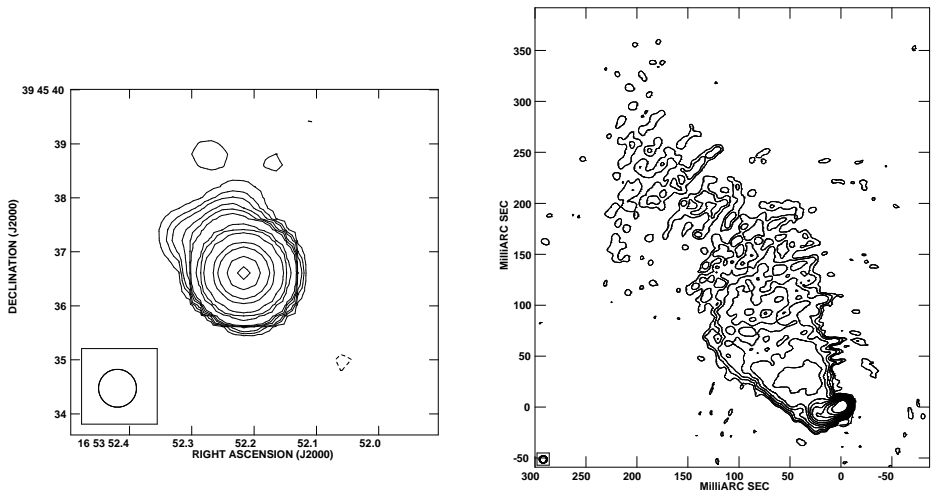}{gg2.ps}

\caption{left: VLA image of Mrk 501. Contours are -1, 1 2 3 5 7 10 30 50 100
300 500 1000 1500 mJy/b; right: HSA image of Mrk 501. Contours are -1 1 1.4
2 2.8 4 5.6 ... $\times$ 0.25 mJy/b}

\end{figure}

In Fig. 1 left, 
we show the image obtained with the VLA during the HSA observations.
The phased array image is dynamical range limited ($\sim$ 10000:1), and it 
shows a one sided 
emission with a short jet like structure at the same PA as the extended
symmetric
structure. 
From this one sided emission we can derive
constraints on the jet velocity ($\beta$c) and orientation ($\theta$) with 
respect to the line of sight. At 2'' we have $\beta$cos$\theta$ $>$ 0.36
and at 1'' $\beta$cos$\theta$ $>$ 0.63. This result implies that at 0.67 kpc 
(projected) from the core the jet is still at least mildly relativistic 
($\beta$ $>$ 0.63).

\section{Sub-arcsecond scale}

Thanks to the good sensitivity and uv-coverage of the HSA data, we can obtain
images of the one-sided jet up to 300 mas from the core at a resolution of
9 $\times$ 5 mas (Fig. 1 right). Near the core the 
jet is bright and compact and after the well known change in its projected
PA, it shows a large opening angle ($\sim$ 40$^\circ$) and a diffuse emission 
which becomes limb-brightened in lower resolution images in agreement with 
previous results \citep{gir04a}.

From the jet brightness assuming intrinsic symmetry we derive that
at 100 mas from the core $\beta$cos$\theta$ $>$ 0.61 and that
at 50 mas from the core $\beta$cos$\theta$ $>$ 0.77

\section {Milliarcsecond scale}

New HSA observations confirm the structure near the core discussed in 
\citet{gir04a}.
The jet in the inner 25 mas is limb-brightned and oriented in PA 110$^\circ$.
With respect to previous results the low noise allows stronger constraints
such that at 10 
mas from the core we derive $\beta$ cos$\theta$ $>$ 0.92.

\section {Sub-milliarcsecond scale}

The 86 GHz VLBI image is shown in Fig. 2 left. 
The HPBW is 0.21 $\times$ 0.09 mas
in PA -3$^\circ$ corresponding to a linear HPBW size of 014 $\times$ 0.06 pc at
the distance of Mrk 501; the image peak flux is 208.2 mJy/b. 
The image shows a bright source at the core position 
with a short one-sided emission in PA 160$^\circ$. At about 1 mas there is
a faint radio knot extended in the E-W direction. Comparing this image with the
22 VLBA GHz image published in \citet{gir04a} 
we can identify this knot as the C4 
region. The brightness decrease from the ``core'' region is very abrupt. No
bright jet is visible despite of the high Doppler factor expected in a
relativistic jet oriented at a small angle with respect to the line of sight.

\begin{figure}[!ht]

\plottwo{gg3.ps}{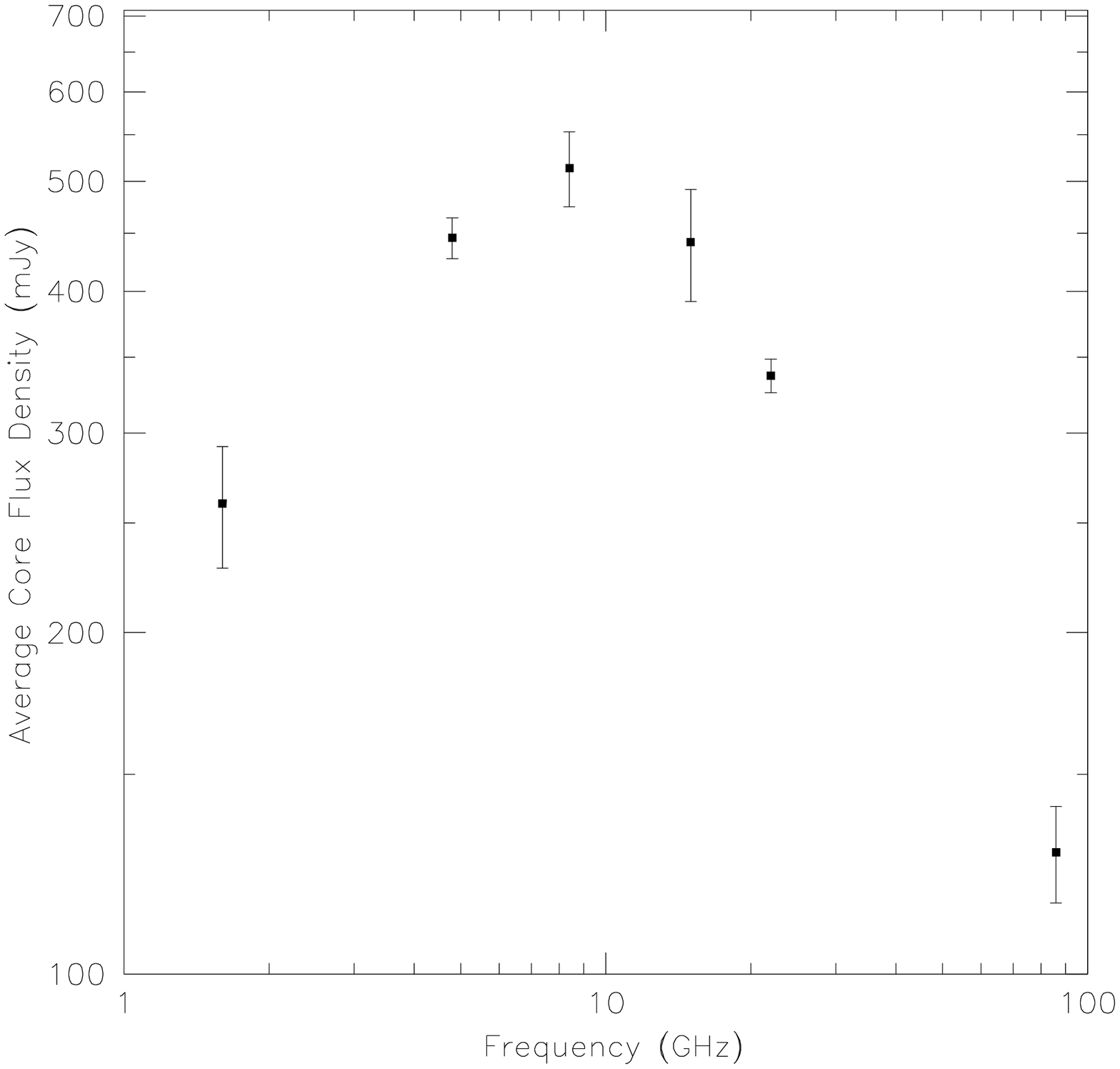}

\caption{left: GMVA image of Mrk 501. Contours are: -1 1 2 4 8 16 32 64 
$\times$ 2 mJy/b (3 $\sigma$ level); right: Radio spectrum of the nuclear 
source of Mrk 501.}

\end{figure}

\section{Discussion}

\begin{itemize}

\item
We estimated the core radio spectral index using the best available measures
\citep{gir04a}. The radio spectrum is shown in Fig. 2 right (note that the
different measurements were obtained at different epochs, however no 
strong radio
variability has been found for Mrk 501). It is self-absorbed at about 8 GHz,
suggesting that observations at higher resolution should resolve the
nuclear source present in the 86 GHz image. The estimated magnetic field
from the self-absorbed spectrum is in the range 0.01 - 0.03 G.

\item
A comparison of the 86 GHz image and the 22 GHz VLBA image \citep{gir04a}
confirms the changes in the projected jet PA:
it is $\sim$ 160$^\circ$ in the inner 1 mas, $\sim$ 90$^\circ$ from 2 to 10 
mas, 110$^\circ$ from 10 to 30 mas where the jet bends to $\sim$ 30$^\circ$ 
afterwhich the jet PA is constant up to the kpc scale.

\item
The Gamma-ray emission detected from Mrk 501 requires a inner high velocity 
jet ($\Gamma$ $\sim$ 15) 
and a small angle with respect to the line of sight ($\theta$ $\sim$ 
4 $^\circ$) \citep{gir04a}. Moreover \citet{ghi05} discuss how a velocity 
structure in the jet could produce Gamma-ray emission.

The absence of a bright jet in the sub-mas image at 86 GHz and the 
limb-brightened structure of the radio jet can be explained assuming a fast 
inner spine and a slower shear layer, with the inner spine at a larger 
angle with respect to the line-of-sight (10 - 15 degree).  This results in 
a higher Doppler factor for the slow shear
layer compared to the fast spine. The reason for the change in the jet PA
with respect to the line-of-sight as well as of the projected jet direction,
is unknown.

\item
The jet velocity decrease is slow: from HSA data we estimate that at 1 
arcsecond (projected) from the core the velocity is still $\beta$ $>$ 0.6. 
This result is
in agreement with the constant jet opening angle ($\sim$ 40$^\circ$) in the
region after the last major change in the jet PA up to $\sim$ 150 mas from the 
core. Such an observed opening angle implies in a freely expanding jet a
Lorentz factor of $\sim$ 5.
At 5 arcsecond from the core the jet is no longer relativistic.

\item
The observational data (in particular the high jet velocity in the inner 
region  and the slow velocity decrease) are in agreement with the adiabatic 
model discussed in \citet{bau97} and in \citet{gir04a}, 
with a magnetic field
mostly perpendicular to the jet direction. A parallel magnetic field predicts
a large velocity decrease in the region imaged using HSA data, in contrast with
the observed results (a counter-jet should be visible well above our detection 
limit). Constraints from HSA and VLA data suggest a jet orientation of $\sim$ 
15$^\circ$ -- 20$^\circ$.

\end{itemize}

\acknowledgements The National Radio Astronomy Observatory (NRAO) is a facility
of the National Science Foundation, operated under cooperative agreement by
Associated Universities, Inc.
We would like to thank Travis A. Rector and all the 
Scientific and Local Organizing Commitee for a job well done. This work was 
partially supported by the Italian Ministry for University and Research
(MIUR) and by the National Institute for Astrophysics (INAF).

\end{document}